\documentclass[aps,amsmath,amssymb,superscriptaddress,preprint]{revtex4-1}
\pdfoutput=1
\usepackage{graphicx}
\usepackage{dcolumn}
\usepackage{bm}
\usepackage{graphicx,color}
\usepackage{epstopdf}
\usepackage{appendix}
\usepackage{ulem}
\begin{document}


\title{Electrical generation and propagation of spin waves in antiferromagnetic thin-film nanostrips}

\author{Xinyi Xu}
\affiliation{Department of Electrical and Computer Engineering, North Carolina State University, Raleigh, NC 27695, USA}
\author{Yuriy G. Semenov}
\affiliation{Department of Electrical and Computer Engineering, North Carolina State University, Raleigh, NC 27695, USA}
\author{Ki Wook Kim}\email{kwk@ncsu.edu}
\affiliation{Department of Electrical and Computer Engineering, North Carolina State University, Raleigh, NC 27695, USA}
\affiliation{Department of Physics, North Carolina State University, Raleigh, NC 27695, USA}


\begin{abstract}
Electrical generation of THz spin waves is theoretically explored in an antiferromangetic nanostrip via the current-induced spin-orbit torque.  The analysis based on micromagnetic simulations clearly illustrates that the N\'{e}el-vector oscillations excited at one end of the magnetic strip can propagate in the form of a traveling wave when the nanostrip axis aligns with the magnetic easy-axis.  A sizable threshold is observed in the driving current density or the torque to overcome the unfavorable anisotropy as expected.  The generated spin waves are found to travel over a long distance while the angle of rotation undergoes continuous decay in the presence of non-zero damping.  The oscillation frequency is tunable via the strength of the spin-orbit torque, reaching the THz regime.  Other key characteristics of the spin waves such as the phase and the chirality can also be modulated actively.  The simulation results further indicate the possibility of wave-like superposition between the excited spin oscillations, illustrating its application as an efficient source of spin-wave signals for information processing.
\end{abstract}
\maketitle


The magnetization fluctuations in the magnetically ordered materials can propagate in the form of a wave, which is the so-called spin wave. From the equivalent quasi-particle point of view, the quanta of spin waves are known as magnons. The spin waves can carry information without involving the charge current flow and, thus, potentially offer an effective medium for  low-power computing technologies.~\cite{Chumak2015,Kruglyak}
A majority of the investigation on the spin waves and their applications have so far been focused on the ferromagnetic (FM) structures. Accordingly, the typical characteristic frequencies of the magnonic devices have been in the GHz range due to the limitation in the intrinsic FM resonance. Recently, the antiferromagnetic (AFM) materials have begun to receive much attention as an alternative for they can in principle exhibit a similar spintronic effect with a much higher response frequency and energy efficiency.~\cite{Jungwirth2016, Cheng2016}
Not only the wave frequency defines the maximum clock rate of a computing device but also a host of other applications beyond the logic/memory realm become accessible once the spin wave reaches the THz regime. The presence of two degenerate modes with opposite circular polarizations in the AFM spin waves offers yet another degree of freedom to encode the information.~\cite{Cheng2016sr, Keffer1952, Keffer1953} The fact that the AFM thin films are not subject to the demagnetization field also makes them a promising candidate for the spin-wave channel as the collective excitations can remain stable over a longer distance. The relative abundance of dielectric antiferromagnets provides an additional flexibility in selecting the materials with desired properties.

The spin-wave excitation has been achieved optically,~\cite{Lenk} thermally,~\cite{Seki} and via electrical methods utilizing the antennas.  In the latter (which is the most commonly adopted approach), the electromagnetic signal applied to a microwave antenna excites magnetization precession in the magnetic material via an alternating Oersted field induced around the antenna.~\cite{Jamali, Demidov, Schneider, Cherepov}  As such, an external microwave source is needed for this scheme, not to mention the lack of electrical tunability. In contrast, a promising alternative approach demonstrated recently~\cite{Demidov2016} took advantage of the magnetization auto-oscillations induced by the spin-transfer torque $-$ a topic of intense investigation since its introduction~\cite{Slonczewski} and guided their propagation along a FM strip.  Since persistent rotations of the N\'{e}el vectors can be excited likewise in the AFM structures via the effective spin torques as shown experimentally in the literature,~\cite{Moriyama2018} it is not unreasonable to anticipate the generation and subsequent propagation of traveling AFM spin waves in a thin-film strip through electrically controlled mechanisms. The approach based on the spin-orbit torque (SOT) is particularly attractive as it can be applied to the broadly available insulating AFM materials.~\cite{Cheng2016}

The studies on the AFM spin-Hall oscillators (i.e., driven by the SOT) have mainly considered the magnetic materials with easy-plane anisotropy, in which the easy plane is aligned normal to the polarization of the induced spin-Hall current.~\cite{Khymyn2017}   As there is no anisotropy to overcome, the N\'{e}el-vector precessions with 90$^\circ$ axial tilt (i.e., right-angle oscillations) can be excited without the threshold.~\cite{Khymyn2017,Gomonay2010}  In other words, even a low driving torque (or the current) can generate the spin oscillations. However, it was found from a theoretical analysis that this configuration is actually not conducive to the traveling waves.~\cite{Semenov2017} When the AFM layer is extended into a long nanostrip waveguide along the axis of rotation,  the Gilbert damping progressively reduces the rotation frequency at the wave front, limiting the number of allowed spiral oscillations away from the excitation point.  The precessions around the hard axis are clearly not the characteristic magnon modes of the AFM material. Thus, a different configuration may be needed for the spin excitations that can subsequently travel a long distance once generated. Note that spin transport studied in context of superfluidity~\cite{Takei2014,Yuan2018,Sonin2019} is different from those excited by the effective spin torques.  Spin superfluid appears in the material with a spectrum of elementary excitations that can transfer spin accumulation. Thus, superfluidity is attributed to the group of magnetic materials with a specific symmetry. By contrast, application of a spin torque creates a magnetic texture that locally breaks the symmetry of the magnetic ordering. In addition, the torque-inducing spin current for the latter is usually applied normal to the intended direction of spin oscillation propagation (e.g., $z$ vs.\ $x$), whereas the spin current injection in the case of superfluidic transport is aligned along the channel axis.

In this work, we theoretically demonstrate that  electrical generation and propagation of N\'{e}el-vector oscillations  can be achieved in an easy-axis AFM thin film. The results from the micromagnetic simulations also illustrate the possibility to actively modulate the key characteristics of the excited spin waves, such as the frequency and the velocity, as well as the superposition between them. This approach utilizing the effective spin torque is expected to be far more efficient and compact than the conventional antenna based method, not to mention the accompanying tunability.

The magnonic structure under consideration is plotted in Fig.~1.  An electrical current flowing through a material with strong spin-orbit coupling [in this case, a heavy metal (HM)] induces the anti-damping SOT in the adjacent AFM layer (with easy $x$-axis). With only a portion of the magnet excited, the problem is different from those studied earlier where the entire nanoparticle was driven by the applied spin torque.~\cite{Cheng2016,Khymyn2017} In addition, the rotations are now around the easy axis ($x$) unlike the cases such as spin superfluidity discussed above (i.e., the hard axis). Both the A-type and G-type AFM materials (preferably insulating or dielectric to avoid the electrical current dissipation) can be used; an A-type is assumed in the current study to be specific.  With the spin oscillations driven directly at one end of the AFM strip, the excited waves are naturally guided along the magnetic layer.  While not shown, additional gate structures can be applied to the channel for further manipulation of the spin-wave properties. Conversion back to the electrical signal can be achieved at the other end of the AFM strip by way of the inverse spin-Hall effect.~\cite{Seki, Chumak2012} The AFM film is also taken to be sufficiently thin in the $z$ direction to ensure the uniformity along this axis (thinner than the typical AFM exchange length of tens of nanometers).~\cite{Abo}


The desired dynamics are analyzed by numerically solving the Landau-Lifshitz-Gilbert (LLG) equation based on Object Oriented MicroMagnetic Framework (OOMMF).~\cite{oommf}  In this approach, the magnetic layer under consideration
is discretized into small cells.  The evolution of the magnetization in each cell is governed by the LLG equation
\begin{equation}
\frac{\partial \mathbf{m}}{\partial t} = \gamma\mathbf{m}\times \mathbf{H}_\mathrm{eff}+\alpha\mathbf{m}\times\frac{\partial \mathbf{m}}{\partial t} +\mathbf{T}^\mathrm{sc} , \label{LLG}
\end{equation}
where $\mathbf{m}$ denotes the unit local sublattice magnetization vector, $\gamma$ is the gyro-magnetic ratio, $\alpha$ gives the Gilbert damping constant, and the macroscopic effective field $\mathbf{H}_\mathrm{eff}\propto\frac{\partial H}{\partial \mathbf{m}}$ is obtained from the Hamiltonian $H$ of the cell that accounts for the exchange interaction and the anisotropy energy.  In the problem at hand, sublattices $A$ and $B$ are coupled antiferromagnetically to each other via the exchange interaction, which can be considered with a simple change of sign in the coupling (to negative). The last term on the right-hand side of Eq.~({\ref{LLG}) describes the anti-damping torque induced by the spin current $\mathbf{J}_s$;~\cite{Gomonay2010,oommf}
\begin{equation}
\mathbf{T}^\mathrm{sc} = \frac{\hbar}{\mu_0e}\frac{\gamma |\mathbf{J}_s|}{2dM_s}(\mathbf{m}\times \hat{\mathbf{\sigma}} \times \mathbf{m}). \label{sot}
\end{equation}
Here, $e$ is the electron charge, $\mu_0$ the vacuum permeability, $d$ the thickness of the magnetic layer, $M_s$ the sublattice magnetization, and $\hat{\mathbf{\sigma}}$  the unit vector of the spin-current polarization.  As the net magnetization is essentially zero ($\mathbf{m}_{A} + \mathbf{m}_{B}\approx 0$), it is more convenient to define the normalized N\'{e}el vector $\mathbf{n}$ given as $\mathbf{m}_{A} - \mathbf{m}_{B}$. Example of OOMMF applied to the AFM systems are readily available in the literature.  For instance, an additional discussion of the  model can be found in Ref.~\onlinecite{Li2017}.

In the actual implementation, $\mathbf{J}_s$ in Eq.~(\ref{sot}) that stems from the spin-Hall effect can be further expressed in terms of the driving electrical current $\mathbf{J}_c$ in the HM layer as
\begin{equation}
    \mathbf{J}_s=\theta_\mathrm{sh} \hat{\sigma}\times {\mathbf{J}_c} , \label{sc}
\end{equation}
where $\theta_\mathrm{sh}$ is the spin-Hall angle. The experimentally measured values of  $\theta_\mathrm{sh}$ are typically in the range of $0.012 \sim 0.12$.~{\cite{Wang2014}} In this calculation, it is assumed that the $y$-directional driving current results in $  \mathbf{J}_s \parallel \hat{\mathbf{z}}$ and $\hat{\mathbf{\sigma}} \parallel \hat{\mathbf{x}}$ with $\theta_\mathrm{sh}=0.1$. The SOT (i.e., $\mathbf{T}^\mathrm{sc}$) is applied only in the initial 30 nm of the AFM strip which has the dimensions of 600$ \times$20$\times$1 nm$^3$.  As for the properties of the AFM film, the easy $x$-axis anisotropy of $K_x = 20$ kJ/m$^3$ is adopted along with the exchange stiffness $A_\mathrm{ex}=5$ pJ/m, $M_s=350$ kA/m, and $\alpha=0.004$ unless mentioned explicitly otherwise.  The anisotropy energy is taken from a frequently cited example of the uniaxial AFM materials, MnF$_2$.~\cite{Sharaev2019}}  Other values are also within the typical range for dielectric magnetic materials used in the literature (e.g., Ref.~\onlinecite{Khymyn2017}).



The feasibility of the desired spin-wave generation is showcased in Fig.~1 with a snapshot of the N\'{e}el vector along the AFM strip (i.e., the $z$-component calculated at an instant).  The color coded plot strongly indicates the formation of a uniform wave pattern through the guiding channel in a steady state that is triggered by local excitation of N\'{e}el-vector oscillations via the SOT.  The decay in the "amplitude" (i.e., the color getting dimmer) is also apparent along the propagation direction as in any oscillators subjected to non-zero damping.  The observed decay in the magnitude of $n_z$ is in fact due to the continued reduction in the angle of rotation around the $x$ direction, resulting in the eventual alignment of the N\'{e}el vector along the easy axis [Fig.~1(b)].  Nonetheless, the helical trajectories along the AFM strip maintain a constant wavelength $\lambda$ and frequency $f$, and can travel over a long distance ($> 1$ $\mu$m)~\cite{comm} without losing the phase information. This process simply corresponds to the gradual relaxation of the initial large-angle oscillations into a characteristic (or typical) spin-wave mode around the easy axis, which is not possible when the excited precession is around a magnetically hard direction (i.e., the case of easy-plane anisotropy). Note that the magnetization dynamics here can be analyzed conveniently in the 1+1 space-time ($x$,$t$) coordinate as the nanostrip dimensions in the $y$ and $z$ directions are taken to be sufficiently smaller than the typical AFM exchange length (i.e., little/no variation in $y$ and $z$).~\cite{Abo}  Accordingly, the rotation frequency can be determined locally from the temporal dynamics of the N\'{e}el-vector trajectories projected on the $y$-$z$ plane. As mentioned above, it stays constant independent of the position $x$ once the system reaches a steady state.

The details of the excited spin-wave dynamics in the region of excitation are more clearly illustrated in Fig.~2.
With the application of the SOT at $t = 0$, the spins are driven away from the easy $x$-axis to oscillate around it ($n_x < 1$).  Similar to the discussion in Ref.~\onlinecite{Cheng2016}, the precession angle (i.e., the axial tilt) gradually increases until it reaches 90$^\circ$ (i.e., the right-angle precession) when the damping is not considered.  In the presence of non-zero $\alpha$, however, the anti-damping spin torque that works to grow the axial tilt gets compensated by the interaction with the magnetic moments in the unexcited part of the AFM strip in driving a propagating wave. As a result,  the N\'{e}el vectors do not fully reach the hard $y$-$z$ plane before settling into a stable trajectory [Figs.~2(a,b)]. Due to the absence of demagnetization field, the AFM oscillations are not subject to such nonlinear effects as foldover.~\cite{Stancil} The angle of rotation around the easy axis is determined by the applied current in combination with the anisotropy energy barrier and the relaxation dynamics [also see Figs.~2(c-e)].  Similarly, it appears that a minimum strength in the SOT is required to excite the spin waves by overcoming the hard $y$-$z$ plane anisotropy (i.e., easy $x$-axis).  For the material conditions specified above, the micromagnetic simulation indicates the threshold current density to be approximately $ 2\times10^7$~A/cm$^2$ $-$ not an exceedingly large value.  Interestingly, this number is in agreement with the estimate of $ \sim 1.5\times10^7$~A/cm$^2$ obtained from an analytical expression given in Ref.~\onlinecite{Cheng2016}.  While the latter was derived for the excitation of spin-Hall oscillations (i.e., no consideration of propagation), it provides a useful criterion here as well since the threshold behavior originates from the similar physical effect in both cases.  The analytical expression indicates approximately a square-root dependence of the critical current density on the easy-axis anisotropy.~\cite{Cheng2016}

One advantage of this electrical spin-wave generation method is the possibility to actively tune the spin-wave frequency in a broad range.  When $J_c$ exceeds the threshold, an intuitive expectation is that a stronger anti-damping torque (i.e., a higher $J_c$) would result in not only a larger angle of rotation but also a higher rotation frequency.  The simulation result plotted in Fig.~3(a) indeed illustrate this behavior where the frequency of the spin-wave eigenmode linearly increases with the driving current, reaching the THz range.  The necessary value for $J_c$ can be reduced if a material with a larger $\theta_\mathrm{sh}$ is used for the spin-orbit coupled layer (such as a topological insulator).
For the traveling mode, the group velocity is another significant parameter as it determines the propagation speed. The dispersion relation between the angular frequency $\omega$ ($= 2 \pi f$) and the wavevector $k$ ($= 2\pi/ \lambda$) shows a nearly linear dependence whose slope can be tailored by, among others, the exchange stiffness $A_\mathrm{ex}$ and the sublattice magnetization $M_s$.  Figure~3(b) provides the $\omega$ vs.\ $k$ plot obtained for two different values of $A_\mathrm{ex}$.  The extracted group velocity is $1.4\times10^6$ cm/s and $2.8\times10^6$ cm/s for $A_\mathrm{ex}=5$ pJ/m and $A_\mathrm{ex}=10$ pJ/m, respectively. The simulation also indicates the inverse proportionality of the velocity to $M_s$.  The observed dependence agrees well with the commonly known relation for the magnon velocity $ v_m = \gamma \sqrt{H_\mathrm{ex} A_\mathrm{ex}/M_s }$, in which the exchange field $H_\mathrm{ex}$ can be further expressed as linear in $A_\mathrm{ex}/M_s$.~\cite{Klingler2015,Gomonay2016}


Along with the frequency and propagation velocity, the phase of the spin wave is another controllable property that can play a significant part in the magnon-based devices.~\cite{Chumak2015} The spin waves propagate a long distance in the AFM strip without losing the phase information even in the presence of non-zero damping.  As the spin oscillation dynamics depend sensitively on the magnetic anisotropy, a change induced in the latter is expected to cause a shift in the spin-wave phase.  The studies in the literature have shown that modulation of the anisotropy profile can be achieved with an electrical control via the magneto-electric or magneto-elastic effects.~{\cite{VCMA,Vaz}}  For instance, the $z$-component of the magnetic anisotropy  can be either strengthened or weakened by switching the polarity of the electrical bias field applied in the same direction.  The introduced change in the anisotropy profile alters the total energy of the system and, thus, the effective field $H_\mathrm{eff}$ in Eq.~(1). Our simulation confirms the subsequent impact on the spin-wave phase, illustrating a possibility to dynamically control this property (see Supplemental Material for additional details).

In analogy to the optical devices, the spin-wave polarization can also be harnessed to encode the information.  Our calculations illustrate that the left or right handedness in the excited N\'{e}el-vector rotation can be determined by the direction of the driving current (i.e., $\mathbf{J}_c \parallel +\hat{\mathbf{y}}$ or $-\hat{\mathbf{y}}$, respectively).  This gives a convenient way to encode binary logic states "1" and "0".  Considering the ease of read-out, on the other hand, the linear polarization offers an alternative scheme that may be more amenable for electrical detection than the encoding based on the chirality.~\cite{Cheng2016sr}  Utilizing the wave-like interference concept discussed earlier in the literature,~\cite{Chumak2015} we explore a possible approach to generate linear polarized spin waves through superposition of multiple spin waves. Specifically, a Y-shaped AFM structure with two input terminals is examined as shown in Fig.~4(a). Two excitation currents, $J_1$ and $J_2$, are applied to the HM layer at the respective input terminals. When $|J_1|=|J_2|$ and antiparallel to each other, two excited spin-wave signals have opposite chirality (one left-handed and the other right-handed), the superposition of which are expected to produce a  linearly polarized wave in the output channel.  Micromagnetic simulations show that this is indeed the case.  As illustrated in Fig.~4(b), the $y$ and $z$ components of the N\'{e}el vector rotate in phase once two input spin waves interfere with each other.
As for $n_x$, it oscillates around a value slightly smaller than 1.  The observed behavior is in contrast to the circularly polarized spin-wave dynamics depicted in Fig.~2(a), where $n_y$ and $n_z$ are 90$^\circ$ out of phase.  Figure~4(c) provides the three-dimension trajectory and its projection on the $y$-$z$ plane at a specific location, more clearly revealing the linearly polarized nature of the output wave. The phases of the two input waves need to be controlled to achieve the desired output signal, for which the effects of random thermal motions can complicate the picture at elevated temperatures. The damping constant $\alpha$ is another parameter that needs to be optimized to extend the range of large-angle oscillations.


In summary, the excitation and propagation of spin waves via the current induced SOT in the AFM thin films are theoretically illustrated. The results from the micromagnetic simulations further highlight the feasibility to actively modulate the key characteristics of the spin waves such as the frequency, velocity, phase, and chirality by exploiting the electrical nature of the control.  The SOT-based approach can also take advantage of other wave-like properties such as the superposition of multiple input signals. The instigated scheme is expected to be far more versatile and energy efficiency than the traditional antenna methods in the magnonic device applications. The anticipated advantages become particularly prominent with an extended range if the damping can be kept to the minimum.  AFM insulators such as hematite ($\alpha$-Fe$_2$O$_3$) may offer this opportunity as a recent experimental study indicates.~\cite{Lebrun2018}

\section*{SUPPLEMENTAL MATERIAL}
See Supplemental Material for the video clip on the phase modulation via the anisotropy change.

\begin{acknowledgments}
This work was supported, in part, by US Army Research Office (W911NF-16-1-0472).  The authors would like to thank Xi-Lai Li for helpful discussions.
\end{acknowledgments}

\clearpage

\begin{figure}
\includegraphics[width=12cm]{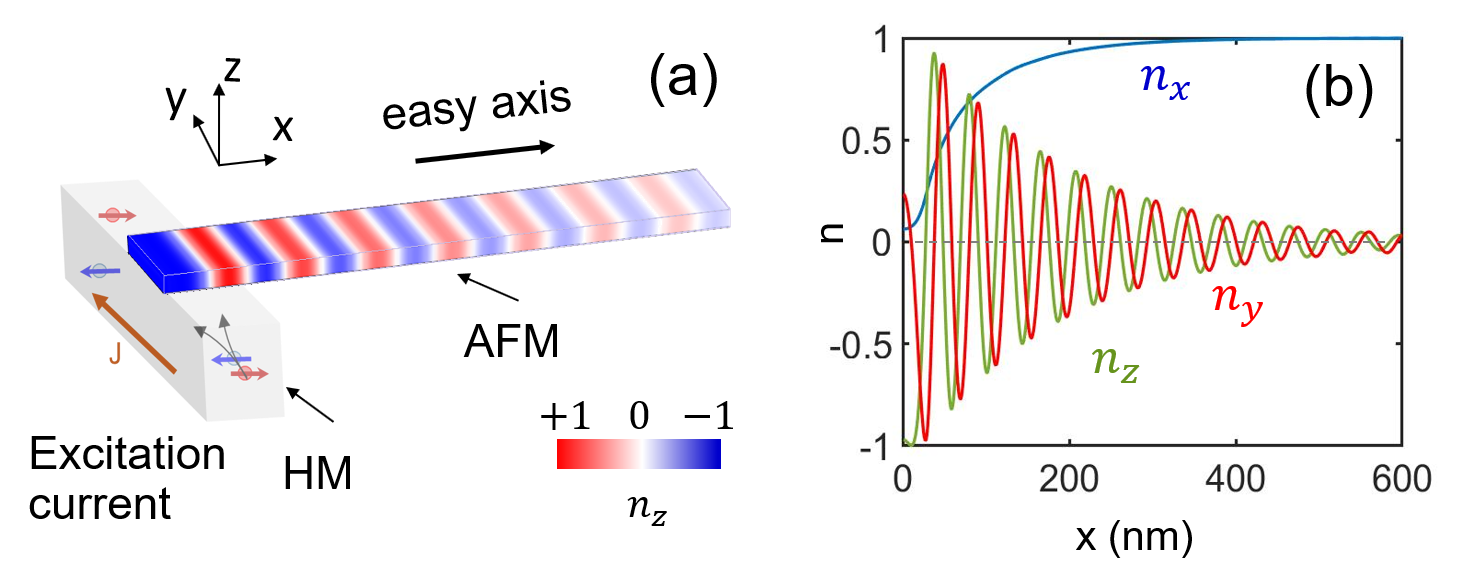}
\caption{(a) Schematic illustration of the spin-wave generation structure under consideration. An electrical current is applied in the HM layer, inducing the spin current and the corresponding SOT in the AFM thin film with easy-axis ($x$) anisotropy.  (b)  A snap shot of the steady-state N\'{e}el vector calculated along the AFM strip. A driving current of $10^8$ A/cm$^2$ is applied for a sufficiently long time to establish a stable spin-wave pattern. To emulate a channel longer than the simulated length (600 nm), a perfectly absorbing boundary is assumed at both ends of the strip (i.e., no reflection). The uniformity of the oscillations in $n_y$ and $n_z$ is clearly visible, indicating a constant frequency and wavelength along the strip.   The color coded plot in (a) illustrates the $z$-component ($n_z$) of this traveling spin wave. Note that the dimensions of the schematic such as the width of the nanostrip are not to the scale for the viewing convenience.}
\end{figure}
\clearpage

\begin{figure}
\includegraphics[width=8cm]{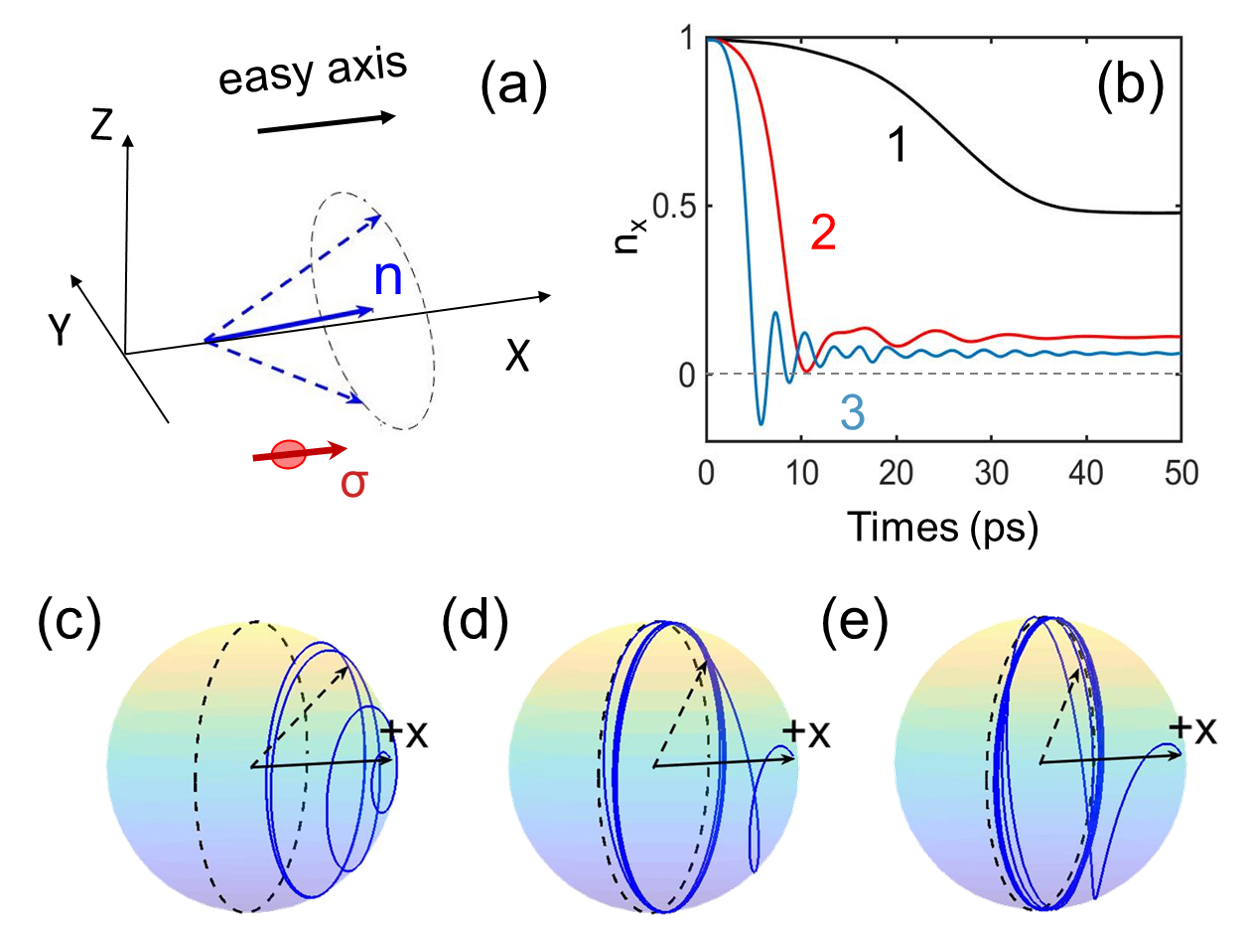}
\caption{(a) Sketch of the N\'{e}el-vector ($\mathbf{n}$) eigen-mode with the polarization ($\mathbf{\sigma}$) of the induced spin-Hall current in the $x$ direction. (b) Temporal evolution of the $x$ component ($n_x$) at the excitation point (i.e., $x=0$) for three driving current densities; $J_c =2\times10^7$~A/cm$^2$ (line 1), $ 5\times10^7$~A/cm$^2$ (line 2), and $ 1\times10^8$~A/cm$^2$ (line 3). (c-e) Corresponding N\'{e}el-vector trajectories in three dimensions (c,d,e - lines l,2,3, respectively).  }
\end{figure}
\clearpage

\begin{figure}
\includegraphics[width=8cm]{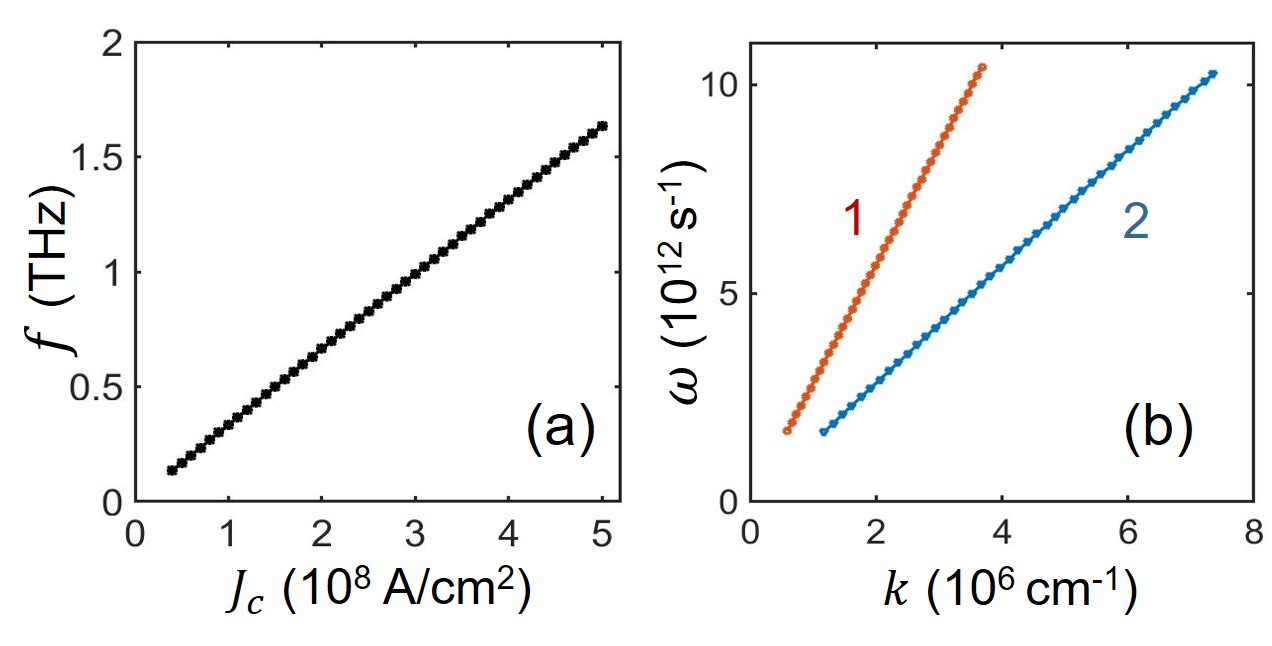}
\caption{(a) Calculated dependence of the AFM spin-wave frequency on the excitation current density $J_c$. (b) Spin-wave dispersion relation with different values of the exchange stiffness constant; $A_\mathrm{ex}=10$ pJ/m (line 1) and $A_\mathrm{ex}=5$ pJ/m (line 2). }
\end{figure}
\clearpage

\begin{figure}
\includegraphics[width=8cm]{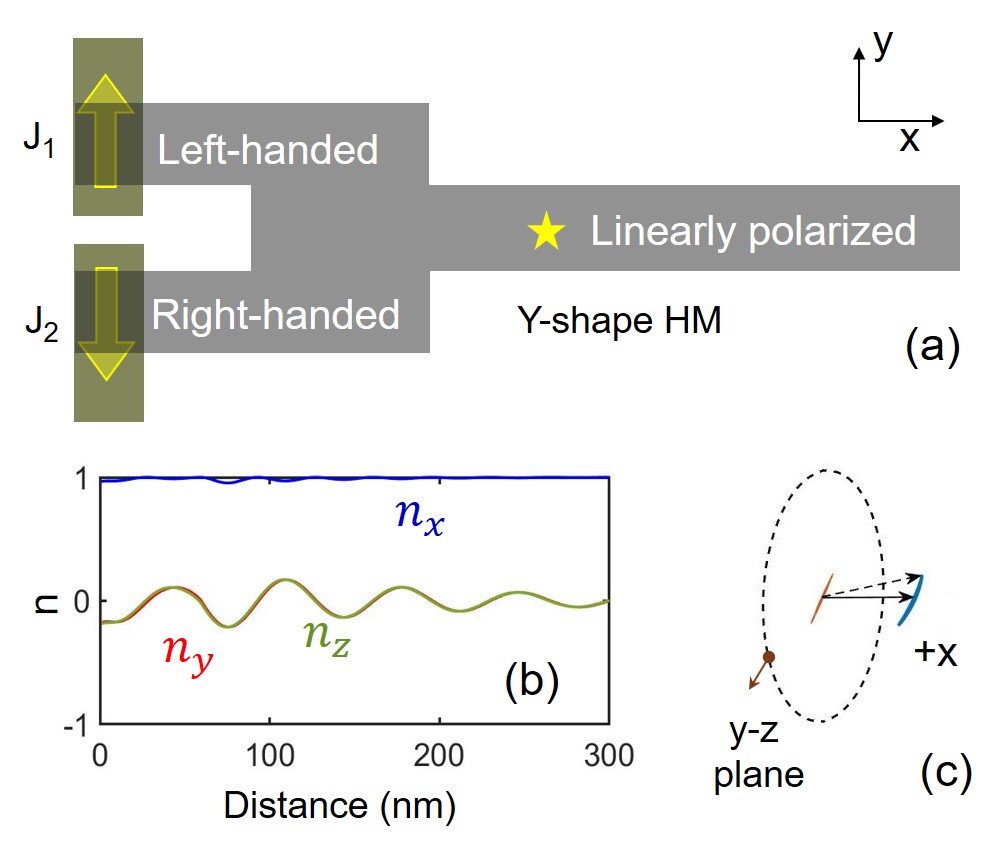}
\caption{(a) Schematic illustration of a Y-shaped AFM structure with two input terminals. As shown, antiparallel currents $J_1$ and $J_2$ excite circularly polarized spin waves of opposite chirality. (b) Calculated N\'{e}el-vector dynamics in the middle channel when two spin waves of opposite chirality interfere with each other. The overlap with the two input terminals occurs over the 60-nm long stretch. (c) Three-dimensional trajectory of the output spin wave and its projection on the $y$-$z$ plane [see the star in (b) for the location in the AFM strip]. The linear polarization is apparent.  In (b,c), the excitation current density of $ 2 \times 10^8$ A/cm$^2$ is used.}
\end{figure}


\clearpage


\begin{references}
\bibitem{Chumak2015} A. V. Chumak, V. I. Vasyuchka, A. A. Serga, and B. Hillebrands, \textit{Nat. Phys.} \textbf{11}, 453 (2015).

\bibitem{Kruglyak} V. V. Kruglyak, S. O. Demokritov, and D. Grundler, \textit{J. Phys. D: Appl. Phys.} \textbf{43}, 264001 (2010).

\bibitem{Jungwirth2016} T. Jungwith, X. Marti, P. Wadley, and J. Wunderlich, \textit{Nat. Nanotechnol.} {\bf 11}, 231 (2016).

\bibitem{Cheng2016} R. Cheng, D. Xiao, and A. Brataas, \textit{Phys. Rev. Lett.} \textbf{116}, 207603 (2016).


\bibitem{Cheng2016sr} R. Cheng, M. W. Daniels, J. G. Zhu, and D. Xiao, \textit{Sci. Rep.} \textbf{6}, 24223 (2016).

\bibitem{Keffer1953} F. Keffer, H. Kaplan, and Y. Yafet, \textit{Am. J. Phys.} \textbf{21}, 250 (1953).

\bibitem{Keffer1952} F. Keffer and C. Kittel, \textit{Phys. Rev.} \textbf{85}, 329 (1952).

\bibitem{Lenk} B. Lenk, H. Ulrichs, F. Garbs, and M. M\"{u}nzenberg, \textit{Phys. Rep.} \textbf{507}, 107 (2011).

\bibitem{Seki} S. Seki, T. Ideue, M. Kubota, Y. Kozuka, R. Takagi, M. Nakamura, Y. Kaneko, M. Kawasaki, and Y. Tokura, \textit{Phys. Rev. Lett.} \textbf{115}, 266601 (2015).

\bibitem{Jamali} M. Jamali, J. H Kwon, S. M. Seo, K. J. Lee, and H. Yang, \textit{Sci. Rep.} \textbf{3}, 3160 (2013).

\bibitem{Schneider} T. Schneider, A. A. Serga, T. Neumann, B. Hillebrands, and M. P. Kostylev, \textit{Phys. Rev. B} \textbf{77}, 214411 (2008).

\bibitem{Demidov} V. E. Demidov, M. P. Kostylev, K. Rott, P. Krzysteczko, G. Reiss, and S. O. Demokritov, \textit{Appl. Phys. Lett.} \textbf{95}, 112509 (2009).

\bibitem{Cherepov} S. Cherepov, P. Khalili Amiri, J. G. Alzate, K. Wong, M. Lewis, P. Upadhyaya, J. Nath, M. Bao, A. Bur, T. Wu, and G. P. Carman, \textit{Appl. Phys. Lett.} \textbf{104}, 082403 (2014).

\bibitem{Slonczewski} J. C. Slonczewski, \textit{J. Magn. Magn. Mater.} \textbf{195}, 261 (1999).

\bibitem{Demidov2016} V. E. Demidov, S. Urazhdin, R. Liu, B. Divinskiy, A. Telegin and S. O. Demokritov, \textit{Nat. Commun.} \textbf{7}, 10446 (2016).

\bibitem{Moriyama2018} T. Moriyama, K. Oda, T. Ohkochi, M. Kimata, and T. Ono, \textit{Sci. Rep.} \textbf{8}, 14167 (2018).


\bibitem{Khymyn2017} R. Khymyn, I. Lisenkov, V. Tiberkevich, B. A. Ivanov, and A. Slavin, \textit{Sci. Rep.} \textbf{7}, 43705 (2017).

\bibitem{Gomonay2010} H. V. Gomonay and V. M. Loktev, \textit{Phys. Rev. B} \textbf{81}, 144427 (2010).

\bibitem{Semenov2017} Y. G. Semenov, X. L. Li, X. Xu, and K. W. Kim, \textit{Phys. Rev. B} \textbf{96}, 224432 (2017).




\bibitem{Takei2014} S. Takei, B. I. Halperin, A. Yacoby, and Y. Tserkovnyak, \textit{Phys. Rev. B} \textbf{90}, 094408 (2014).

\bibitem{Yuan2018} W. Yuan, Q. Zhu, T. Su, Y. Yao, W. Xing, Y. Chen, Y. Ma, X. Lin, J. Shi, R. Shindou, X. C. Xie, W. Han, \textit{Sci. Adv.} \textbf{4}, eaat1098 (2018).

\bibitem{Sonin2019} E. B. Sonin, \textit{Phys. Rev. B} \textbf{99}, 104423 (2019).

\bibitem{Chumak2012} A. V. Chumak, A. A. Serga, M. B. Jungfleisch, R. Neb, D. A. Bozhko, V. S. Tiberkevich, and B. Hillebrands, \textit{Appl. Phys. Lett.} \textbf{100}, 082405 (2012).

\bibitem{Abo} G. S. Abo, Y. Hong, J. Park, J. Lee, W. Lee, and B. Choi, \textit{IEEE Trans. Magn.} \textbf{49}, 4937 (2013).

\bibitem{oommf} M. J. Donahue and D. G. Porter, \textit{OOMMF User's Guide}, Version 1.0 (NIST, Gaithersburg, MD, 1999).

\bibitem{Li2017}
X.-L. Li, X. Duan, Y. G. Semenov, and K. W. Kim, \textit{J. Appl. Phys.} \textbf{121}, 023907 (2017).

\bibitem{Wang2014} Y. Wang, P. Deorani, X. Qiu, J. H. Kwon, and H. Yang, \textit{Appl. Phys. Lett.} \textbf{105}, 152412 (2014).

\bibitem{Sharaev2019}
A. Y. Sharaevskaya, D. V. Kalyabin, E. N. Beginin, Y. K. Fetisov, and S. A. Nikitov, \textit{J. Magn. Magn. Mater.} \textbf{475}, 778 (2019).

\bibitem{comm}
Additional simulations with longer strip lengths ($ > 1~\mu$m) were also used to examine this point (not shown).

\bibitem{Stancil}
D. D. Stancil and A. Prabhakar, \textit{Spin Waves: Theory and Applications} (Springer, New York, 2008).

\bibitem{Klingler2015}
S. Klingler, A. V. Chumak, T. Mewes, B. Khodadadi, C. Mewes, C. Dubs, O. Surzhenko, B. Hillebrands, and A. Conca, \textit{J. Phys. D: Appl. Phys.} \textbf{48}, 015001 (2015).

\bibitem{Gomonay2016}
O. Gomonay, T. Jungwirth, and J. Sinova, \textit{Phys. Rev. Lett.} \textbf{117}, 017202 (2016).

\bibitem{VCMA} P. K. Amiri and K. L. Wang, \textit{Spin} \textbf{2}, 1240002 (2012).

\bibitem{Vaz} C. A. F. Vaz, \textit{J. Phys.: Condens. Matter} \textbf{24}, 333201 (2012).

\bibitem{Lebrun2018}
R. Lebrun, A. Ross, S. A. Bender, A. Qaiumzadeh, L. Baldrati, J. Cramer, A. Brataas, R. A. Duine, and M. Kl\"{a}ui, \textit{Nature} \textbf{561}, 222 (2018).



\end{references}
\end{document}


\title{Supplemental Material: Electrical generation and propagation of spin waves in antiferromagnetic thin-film nanostrips}

\author{Xinyi Xu}
\affiliation{Department of Electrical and Computer Engineering, North Carolina State University, Raleigh, NC 27695, USA}
\author{Yuriy G. Semenov}
\affiliation{Department of Electrical and Computer Engineering, North Carolina State University, Raleigh, NC 27695, USA}
\author{Ki Wook Kim}\email{kwk@ncsu.edu}
\affiliation{Department of Electrical and Computer Engineering, North Carolina State University, Raleigh, NC 27695, USA}
\affiliation{Department of Physics, North Carolina State University, Raleigh, NC 27695, USA}
\maketitle

\section*{S1. Phase Modulation}
\begin{figure}[h]
\includegraphics[width=8cm]{spin-wave_s1.jpg}
\caption{Phase modulation via the magnetic anisotropy change introduced to the AFM structure.  A $z$-directional anisotropy of $40$~kJ/m$^3$ is added to the upper channel (see the boxed region marked PMA), while the spin waves in both thin-film strips are generated at the left end of the device.  A driving current of $10^8$ A/cm$^2$ ($\alpha = 0.001$) is used. (Multimedia view)}
\end{figure}